\documentclass[conference]{IEEEtran}
\IEEEoverridecommandlockouts
\usepackage{cite}
\usepackage{amsmath,amssymb,amsfonts}
\usepackage{algorithmic}
\usepackage{graphicx}
\usepackage{textcomp}
\usepackage{xcolor}
\usepackage{booktabs}
\usepackage{hyperref}
\usepackage{placeins}
\def\BibTeX{{\rm B\kern-.05em{\sc i\kern-.025em b}\kern-.08em
    T\kern-.1667em\lower.7ex\hbox{E}\kern-.125emX}}
\begin{document}

\title{Window Size Versus Accuracy Experiments in Voice Activity Detectors\\
}

\author{\IEEEauthorblockN{Max McKinnon, Samir Khaki, Chandan KA Reddy, William Huang}
\IEEEauthorblockA{\textit{Google LLC} \\
Mountain View, U.S.A. \\
mckinnonm@google.com}
}

\maketitle

\begin{abstract}
Voice activity detection (VAD) plays a vital role in enabling applications such as speech recognition. We analyze the impact of window size on the accuracy of three VAD algorithms: Silero, WebRTC, and Root Mean Square (RMS) across a set of diverse real-world digital audio streams. We additionally explore the use of hysteresis on top of each VAD output. Our results offer practical references for optimizing VAD systems. Silero significantly outperforms WebRTC and RMS, and hysteresis provides a benefit for WebRTC.
\end{abstract}



\section{Introduction}
Voice Activity Detection (VAD) is foundational for speech applications, particularly with gating storage and gating more computationally expensive processing. This work investigates the performance effects of extended time windows across three VADs: Root Mean Square (RMS), WebRTC \cite{b1}, and Silero \cite{b2}.

\section{Related Work}
VADs have a rich history, from energy-based methods \cite{b3, b4} to neural approaches \cite{b5}. Early work in VAD focused on energy-based methods and statistical models \cite{b3, b4}. Receiver Operating Characteristics (ROC) and Area Under Curve (AUC) were developed as standard evaluation in \cite{b6}. WebRTC VAD became a widely used open-source baseline \cite{b1}. A more recently developed Silero VAD offered more a performant baseline \cite{b2}. Postprocessing with hysteresis and state was explored in \cite{b7, b8}. This paper builds upon the existing work by exploring much larger window sizes from 10 milliseconds up to 10 seconds.

\section{Experimental Setup}

\subsection{Dataset}
The internal 1.1-hour test dataset targets digital audio streams and consists of 21 diverse files ranging from podcasts to movies to music to documentaries, representing typical digital audio playback content.

\begin{itemize}
    \item English speech
    \item 16bit PCM wav
    \item 16kHz sampling rate
    \item Single channel

\end{itemize}

\subsection{Evaluated VADs}
The following VADs are used in this study. In addition, each VAD is evaluated with hysteresis tuned via grid search.

\begin{itemize}
    \item RMS: A simple baseline VAD
    \item WebRTC: open, widely deployed VAD
    \item Silero: open, known for performance and robustness
\end{itemize}

For WebRTC and Silero, which have fixed window sizes 10ms and 16ms respectively, larger effective windows are constructed by subdividing each larger window into model-compatible subwindows and averaging the predictions across them.

\subsubsection{RMS VAD Details}

The RMS signal of a 16b audio array is calculated in dBFS and then linearly mapped with outliers clipped so -100 dBFS to 0 dBFS is equivalent to a prediction of $[0.0, 1.0]$.

Let the intermediate prediction score, $p'$, be defined as:
\begin{equation}
  p' = \frac{\mathrm{RMS}_{\mathrm{dBFS}} + 100}{100},
\end{equation}
The final prediction, $p$, is then clipped to the range $[0.0, 1.0]$:
\begin{equation}
  p = \max(0, \min(1, p')).
\end{equation}

\subsection{Metrics}

\begin{itemize}
    \item True Positive Rate (TPR) and False Positive Rate (FPR) ROC AUC
    \item Precision-recall (PR) Average Precision (AP)
    \item Matthews Correlation Coefficient (MCC), across hysteresis thresholds
\end{itemize}

Because of the dataset imbalance reflecting typical distributions of speech and no-speech, two types of AUC measurements are assessed.

While ROC AUC summarizes how well the model separates positives from negatives, regardless of the class balance, it can potentially mask performance issues because even a low FPR means many false positives in absolute numbers. Randomly guessing will always have an ROC AUC of 0.5 regardless of class imbalance.

PR AP focuses on how well the model identifies true positives among the predicted positives.

The relationship between ROC AUC and PR AUC is further discussed in \cite{b10}.

MCC is a metric that is similar to a single number F1 tuning metric but also takes into account all four quadrants of the confusion matrix and is robust even in severe class imbalance. MCC varies between -1 (total disagreement) and +1 (perfect prediction) with 0 representing random performance regardless of class distribution.

\section{Results}

\subsection{Initial Findings}
For a typical 100ms window setting and comparing all three VADs, Silero performed significantly better than WebRTC which performed significantly better than RMS.

\begin{figure}[htbp]
\centerline{\includegraphics[width=1.0\columnwidth]{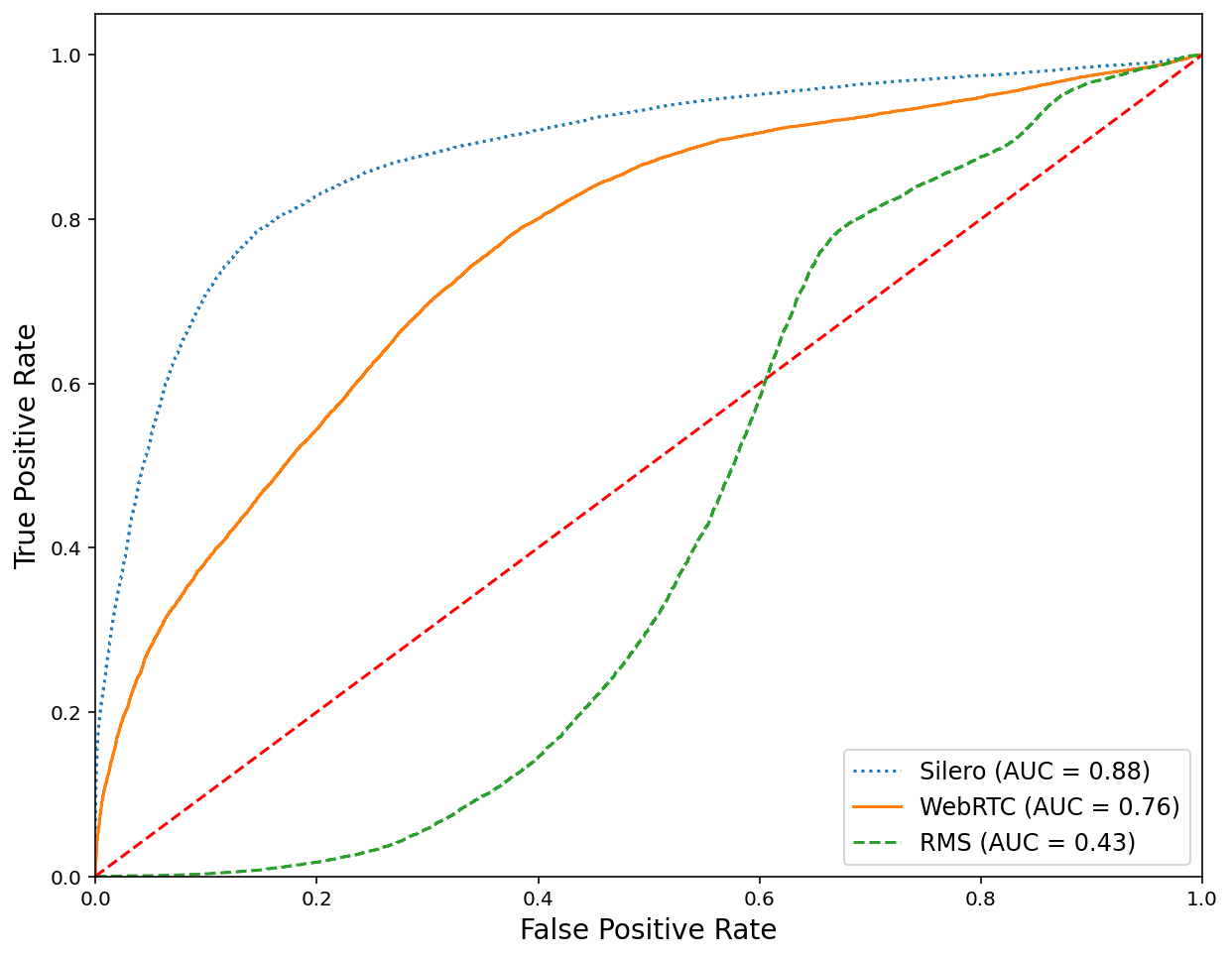}}
\caption{ROC for 100ms windows.}
\label{fig_roc}
\end{figure}

As might be expected from the RMS simple VAD which has no information about the type of signal but just its level, it has a relatively narrow sweet spot and underperforms random guessing for most of its range as seen in Fig.~\ref{fig_roc}.

\begin{figure}[htbp]
\centerline{\includegraphics[width=1.0\columnwidth]{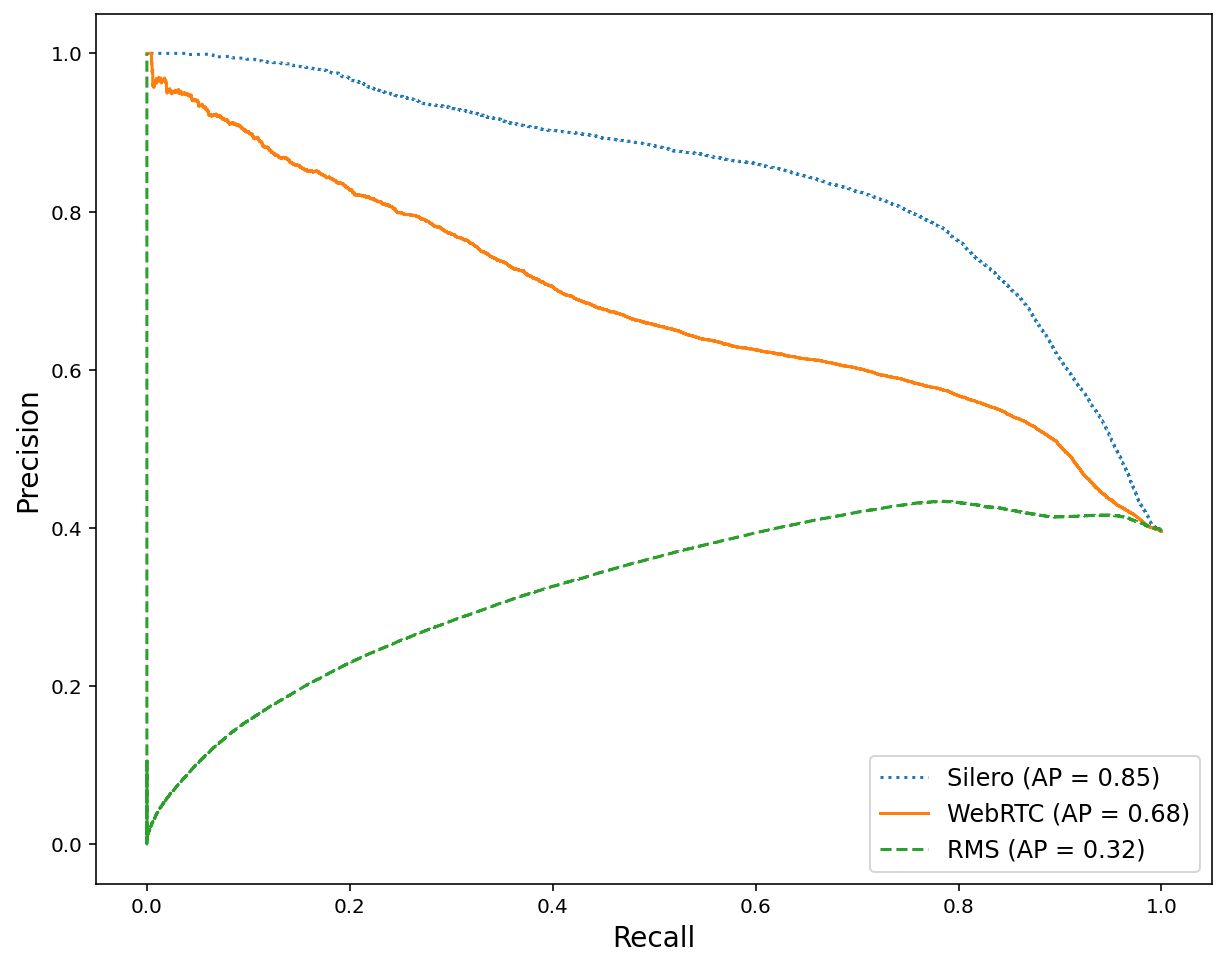}}
\caption{Precision-recall for 100ms windows.}
\label{fig_pr}
\end{figure}

In the PR curve in Fig.~\ref{fig_pr}, we observe about 40\% of the duration of the dataset under test is speech by the precision at a recall of 1.0.

\subsection{Performance vs Window Size}
\begin{figure}[htbp]
\centerline{\includegraphics[width=1.0\columnwidth]{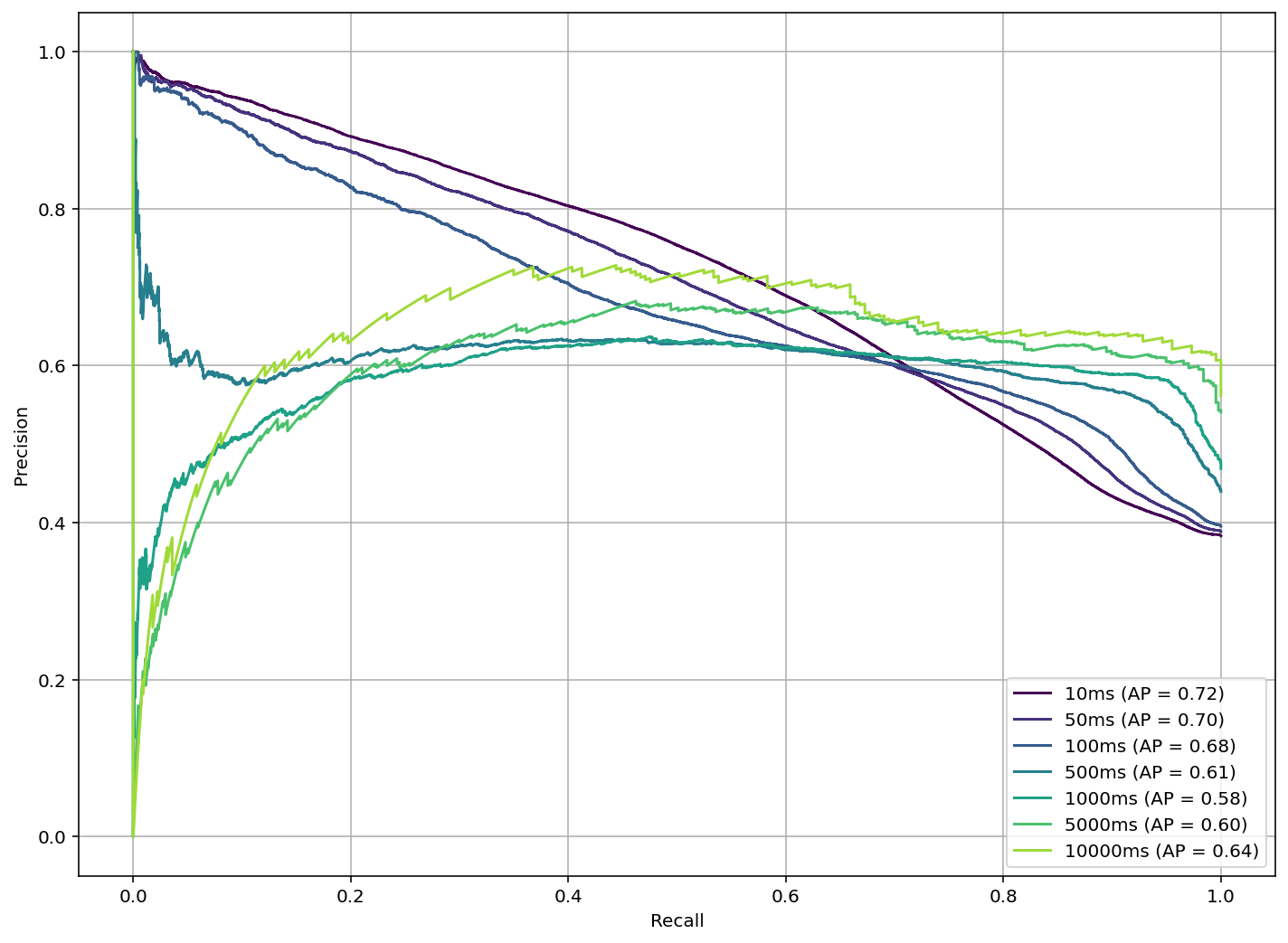}}
\caption{WebRTC precision-recall across window sizes.}
\label{fig_ap_webrtc}
\end{figure}

\begin{figure}[htbp]
\centerline{\includegraphics[width=1.0\columnwidth]{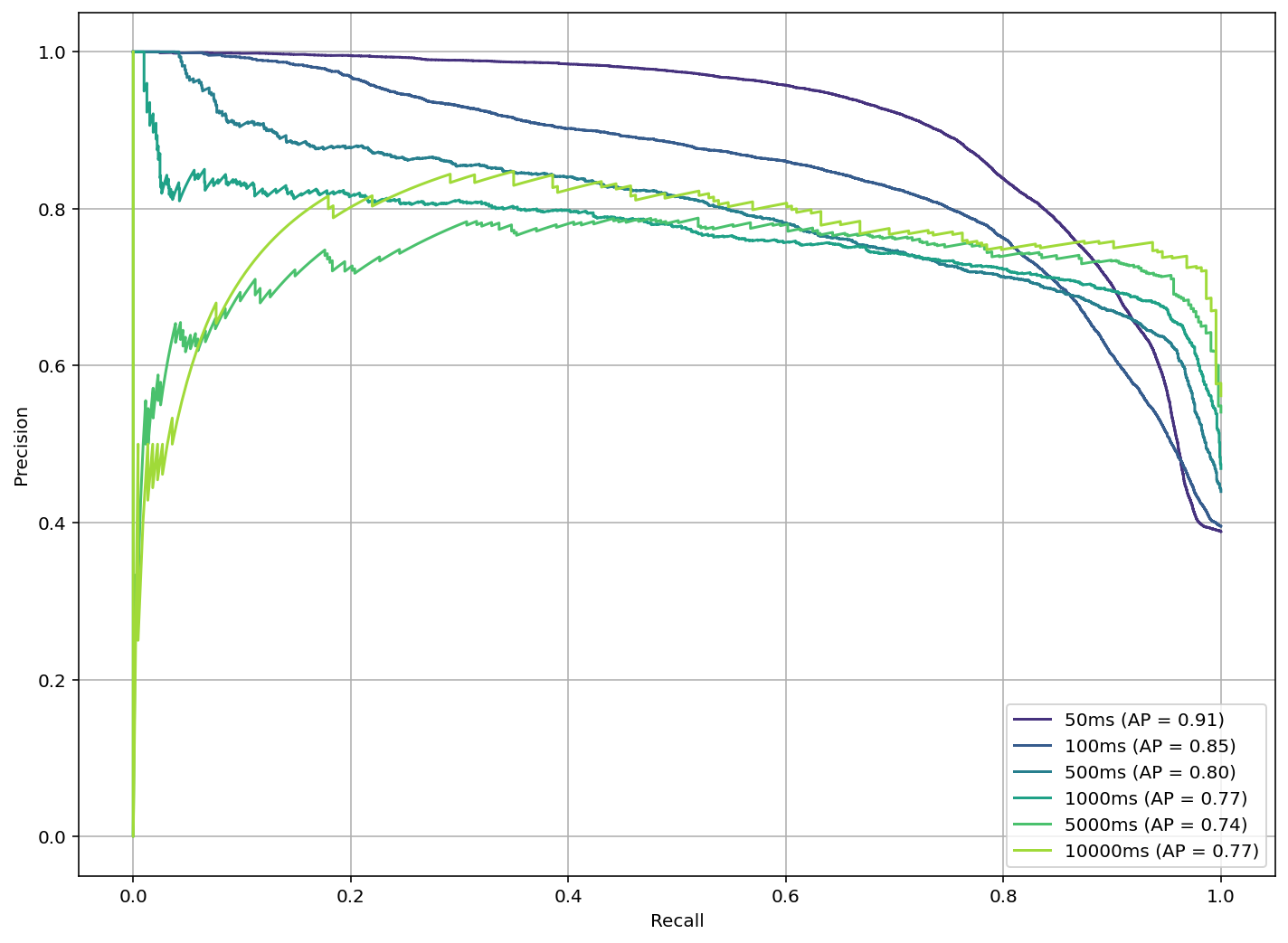}}
\caption{Silero precision-recall across window sizes.}
\label{fig_ap_silero}
\end{figure}

An interesting phenomenon can be seen in the 0.8 to 1.0 recall range in Fig.~\ref{fig_ap_webrtc} and Fig.~\ref{fig_ap_silero} where the bigger windows are actually better than the smaller windows. This is possibly due to the way the ground truth windows are created: if any speech is in the window, it is considered to be a speech window. This is also evident by the traces ending at slightly different points of precision. It could also be partially due to spaces between words and the ground truth of no-speech not labeled granularly in between words and phrases of what would still be considered a continuous speech chunk for most downstream use cases.

\begin{figure}[htbp]
\centerline{\includegraphics[width=1.0\columnwidth]{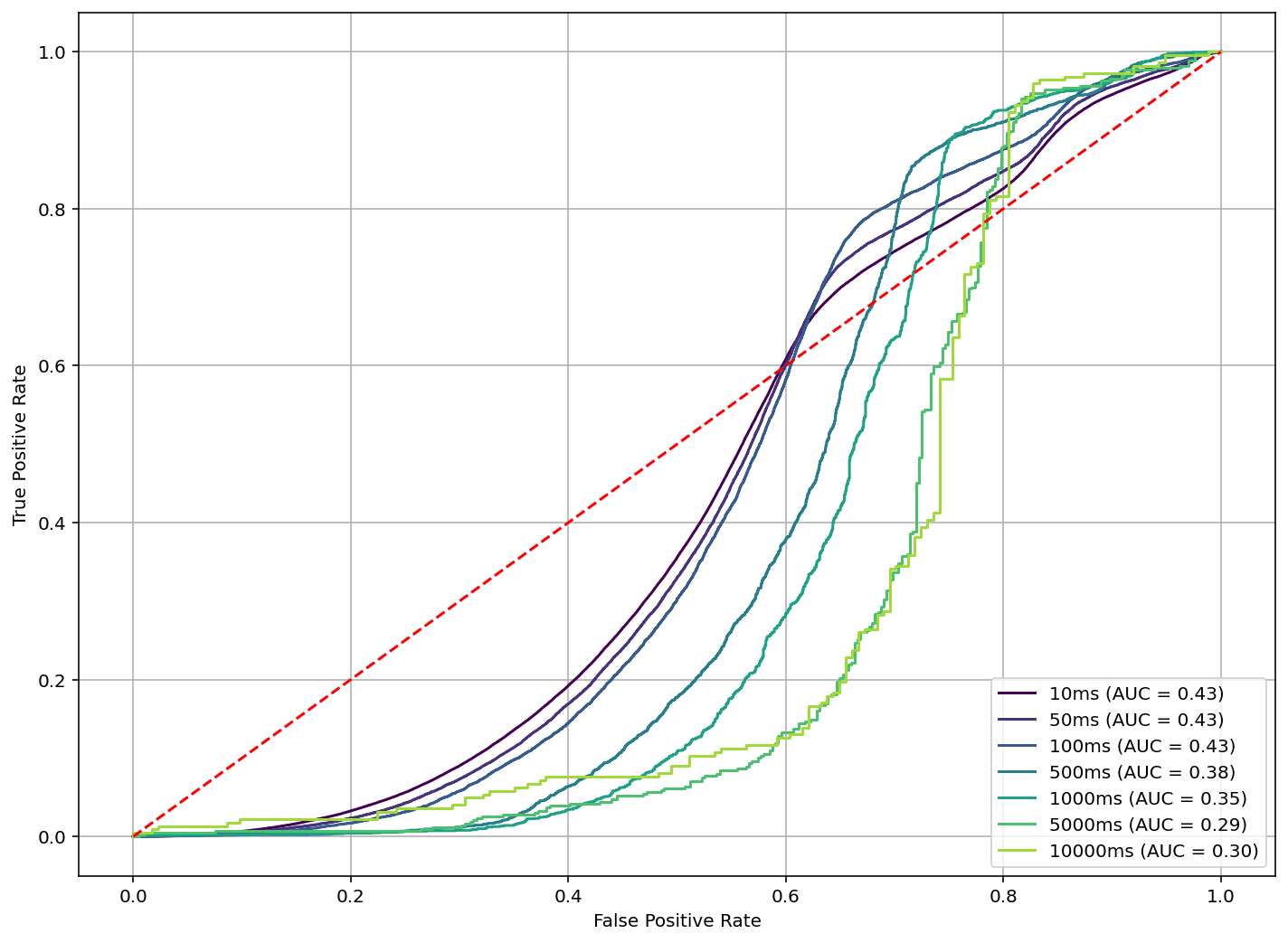}}
\caption{RMS ROC across window sizes.}
\label{fig_auc_rms}
\end{figure}

\begin{figure}[htbp]
\centerline{\includegraphics[width=1.0\columnwidth]{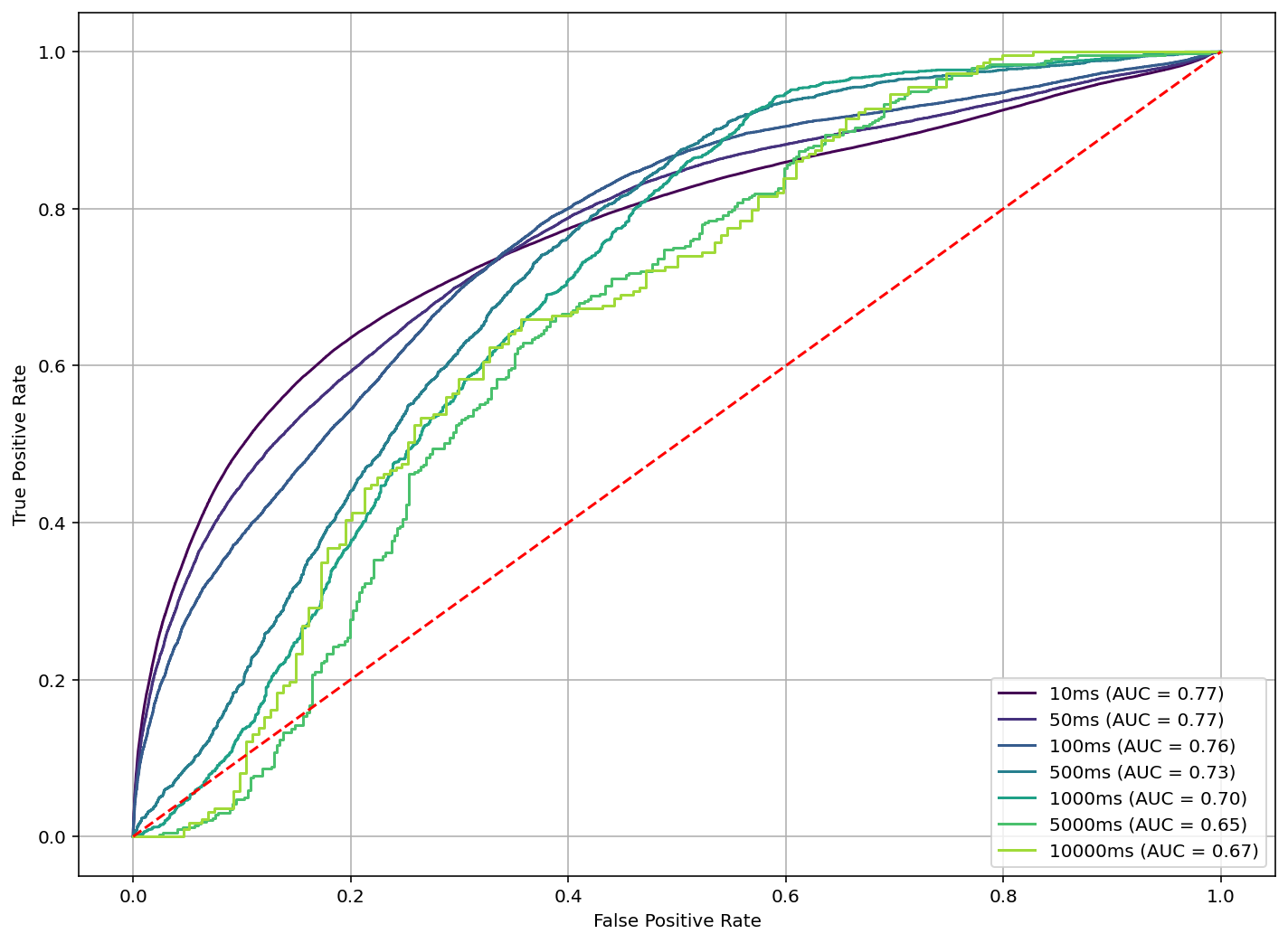}}
\caption{WebRTC ROC across window sizes.}
\label{fig_auc_webrtc}
\end{figure}

\begin{figure}[htbp]
\centerline{\includegraphics[width=1.0\columnwidth]{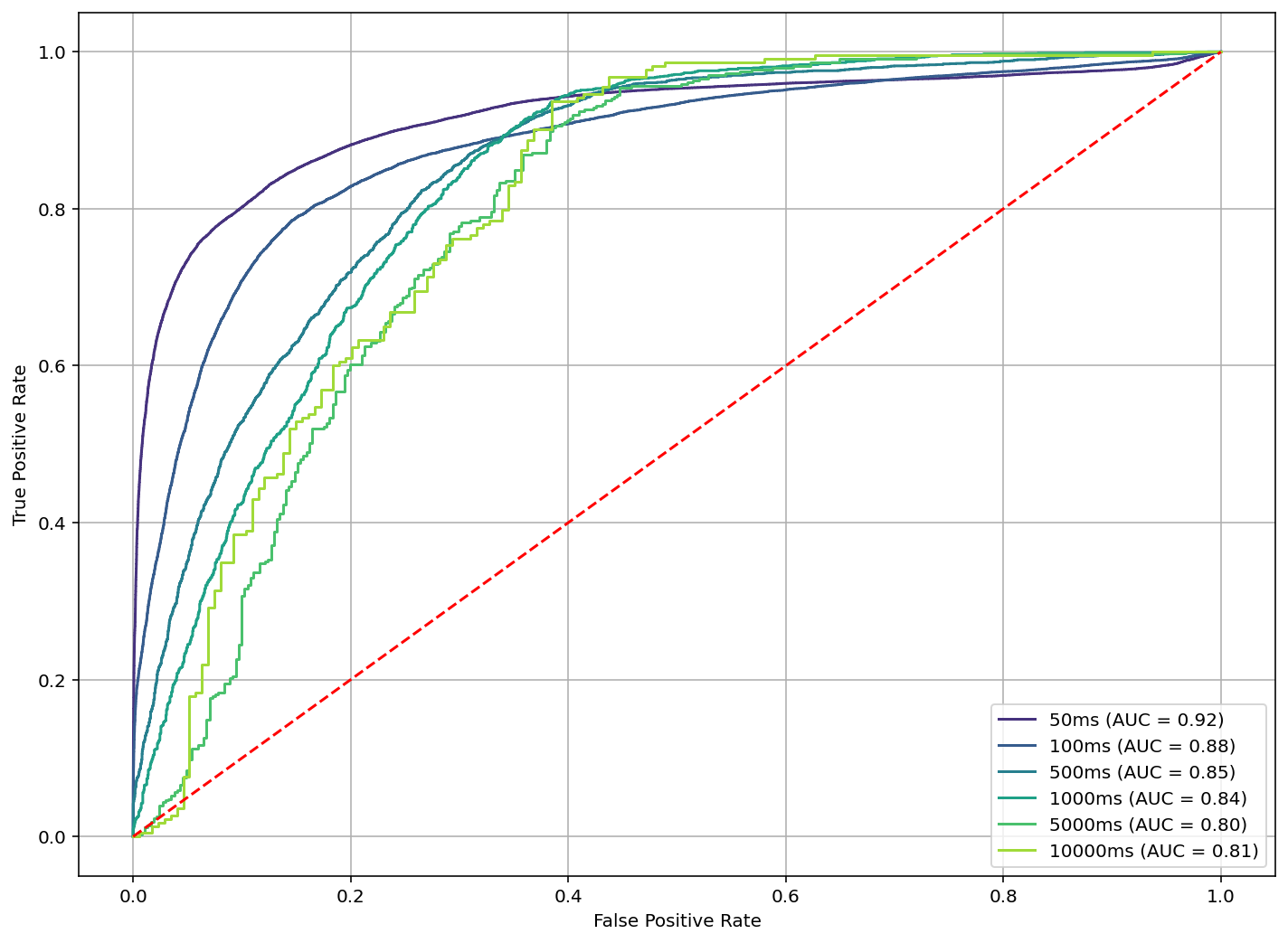}}
\caption{Silero ROC across window sizes.}
\label{fig_auc_silero}
\end{figure}

\begin{figure}[htbp]
\centerline{\includegraphics[width=1.0\columnwidth]{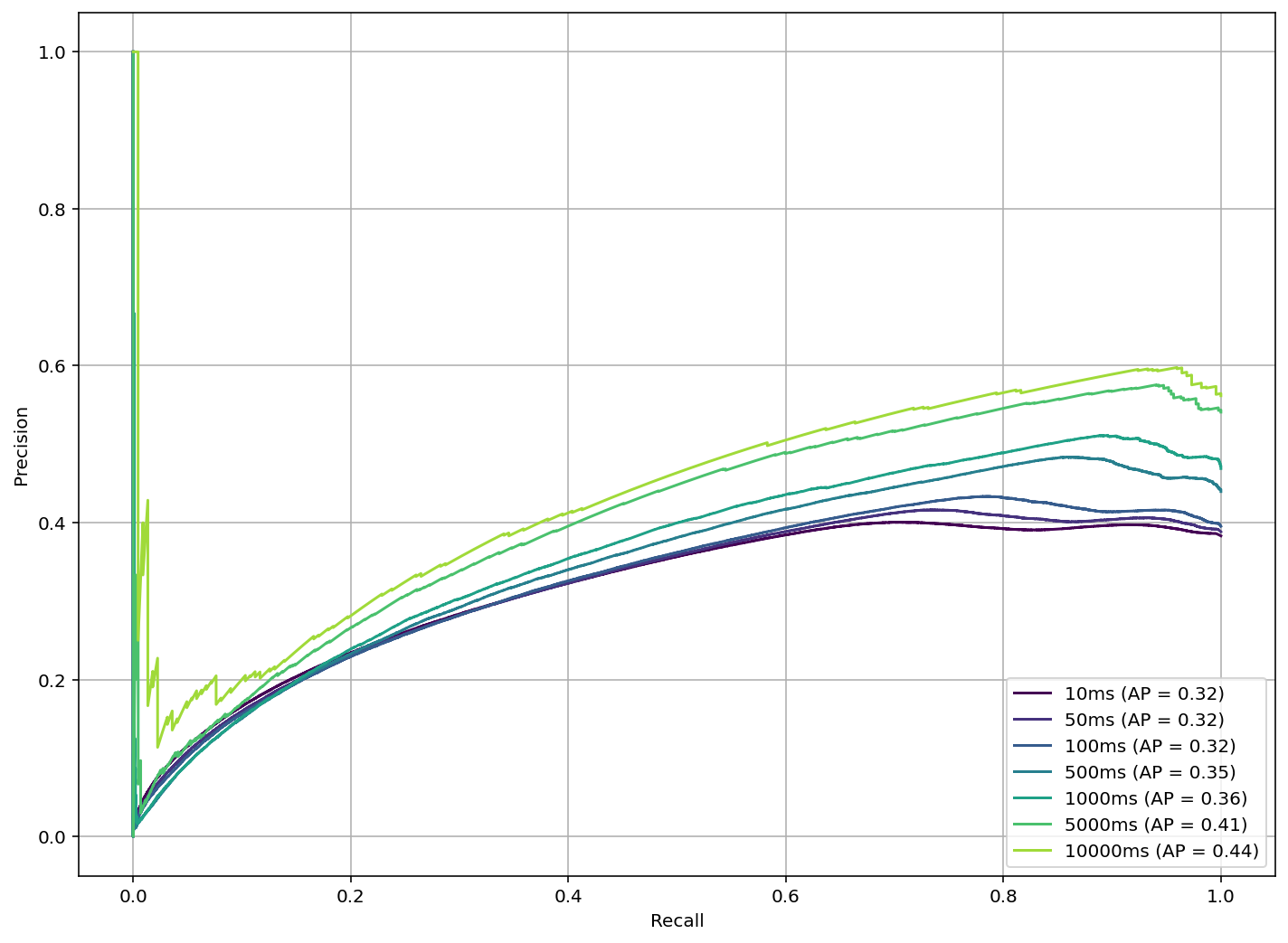}}
\caption{RMS precision-recall across window sizes.}
\label{fig_ap_rms}
\end{figure}

\subsection{Performance Across Dataset}
Some of the data is more challenging than others, including music audio streams seemed to be challenging for all the models to false positive as speech. A distribution of scores by model and window size can be seen in the following figures.

\begin{figure}[htbp]
\centerline{\includegraphics[width=1.0\columnwidth]{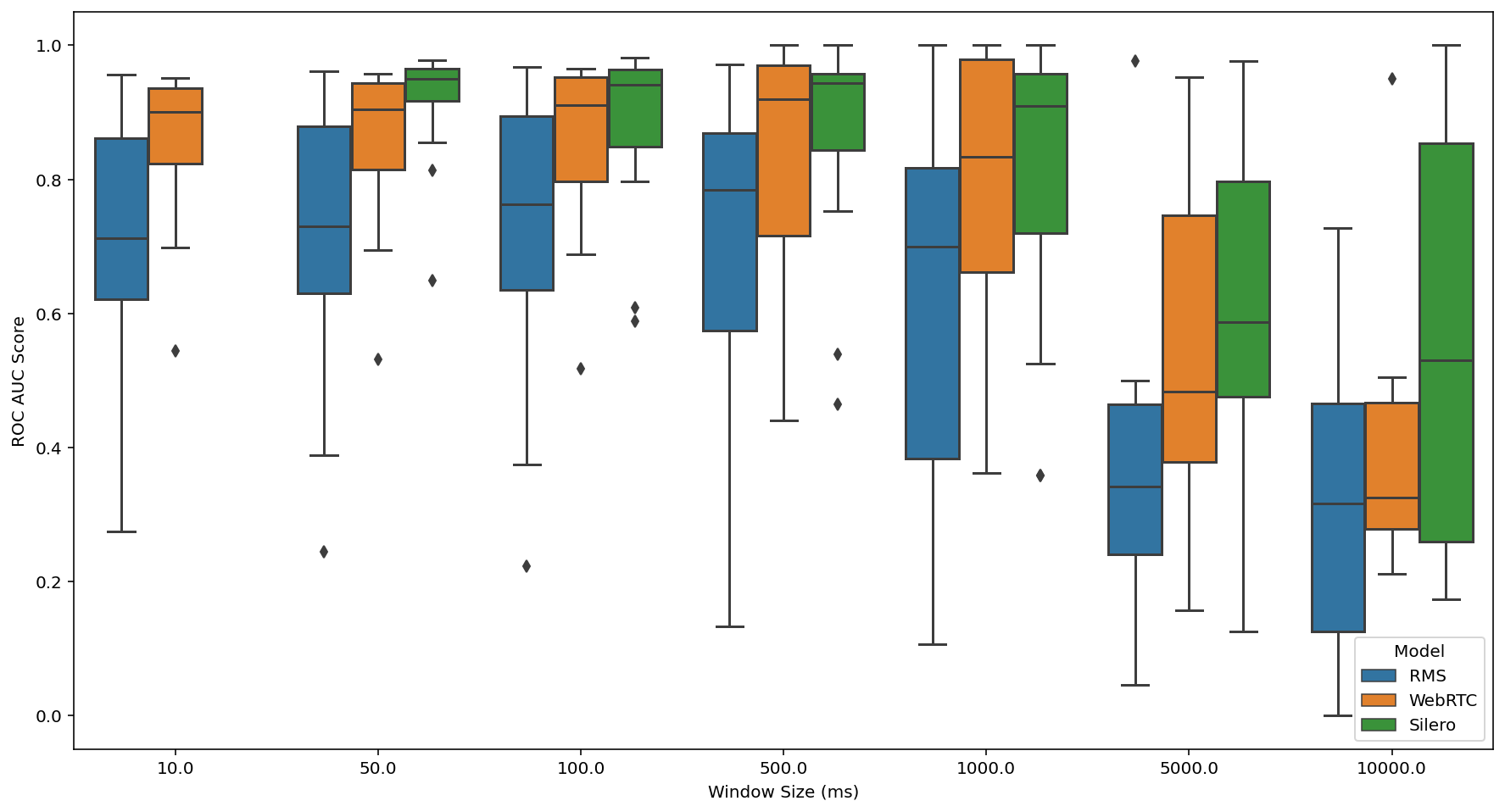}}
\caption{ROC AUC across window sizes per model.}
\label{fig_distribution}
\end{figure}

\subsection{Performance of Adding Hysteresis (50ms Windows)}
Next we introduce a hysteresis state machine around the 50ms version of each classifier. The hysteresis has a high threshold, the threshold to turn on, and a low threshold, the threshold to turn off.

\begin{figure}[htbp]
\centerline{\includegraphics[width=0.9\columnwidth]{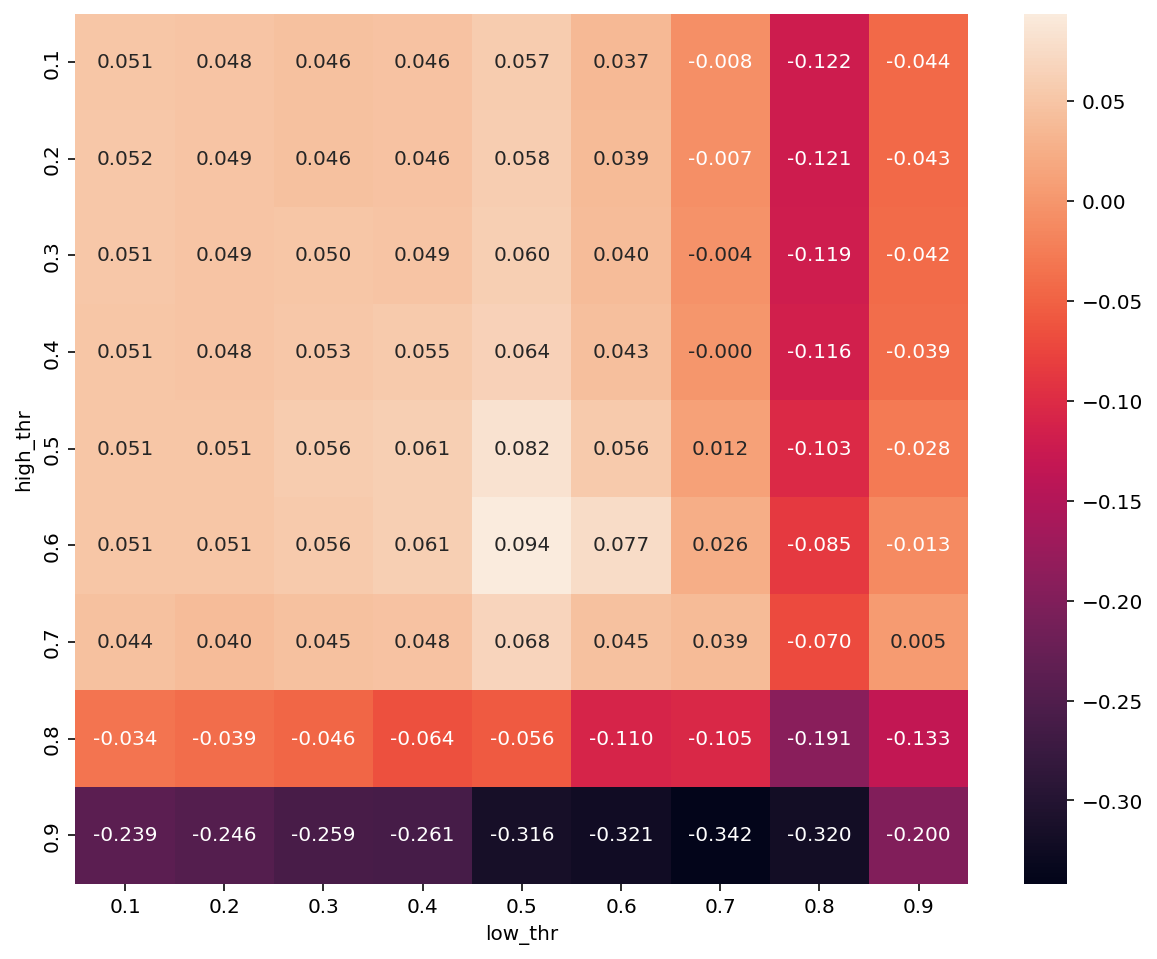}}
\caption{RMS MCC vs hysteresis thresholds.}
\label{fig_mcc_rms}
\end{figure}

\begin{figure}[htbp]
\centerline{\includegraphics[width=0.9\columnwidth]{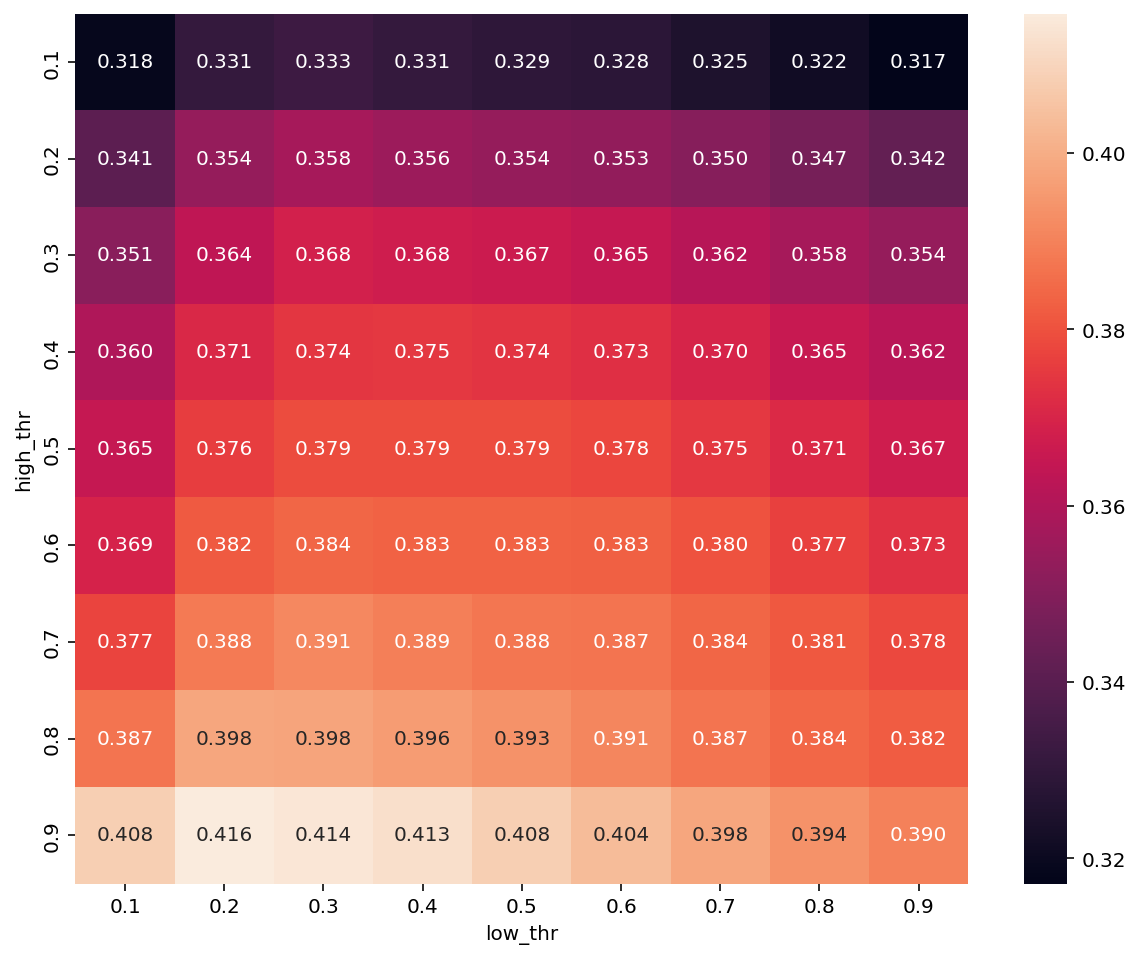}}
\caption{WebRTC MCC vs hysteresis thresholds.}
\label{fig_mcc_webrtc}
\end{figure}

\begin{figure}[htbp]
\centerline{\includegraphics[width=0.9\columnwidth]{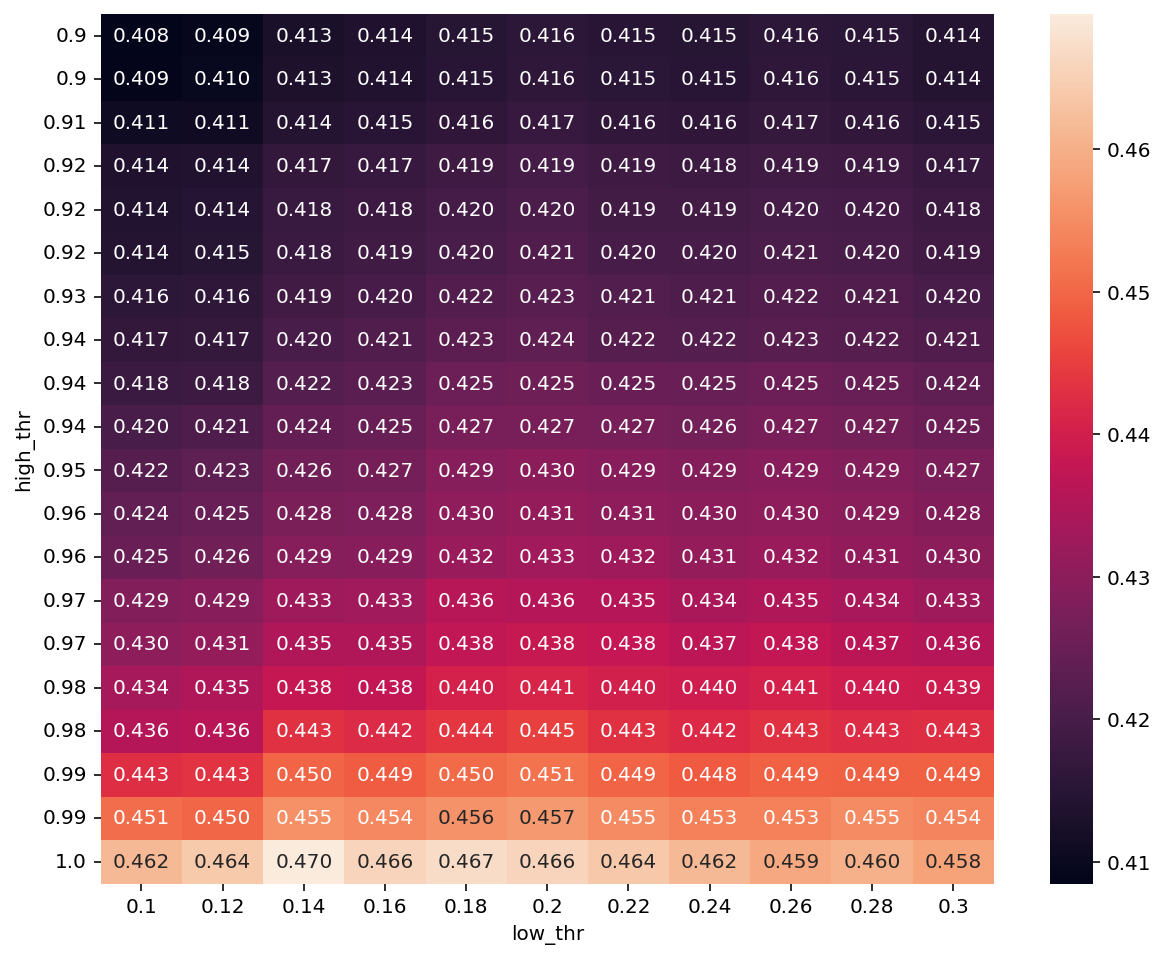}}
\caption{WebRTC MCC vs hysteresis thresholds (zoomed in around optimum).}
\label{fig_mcc_webrtc_zoomed}
\end{figure}

\begin{figure}[htbp]
\centerline{\includegraphics[width=0.9\columnwidth]{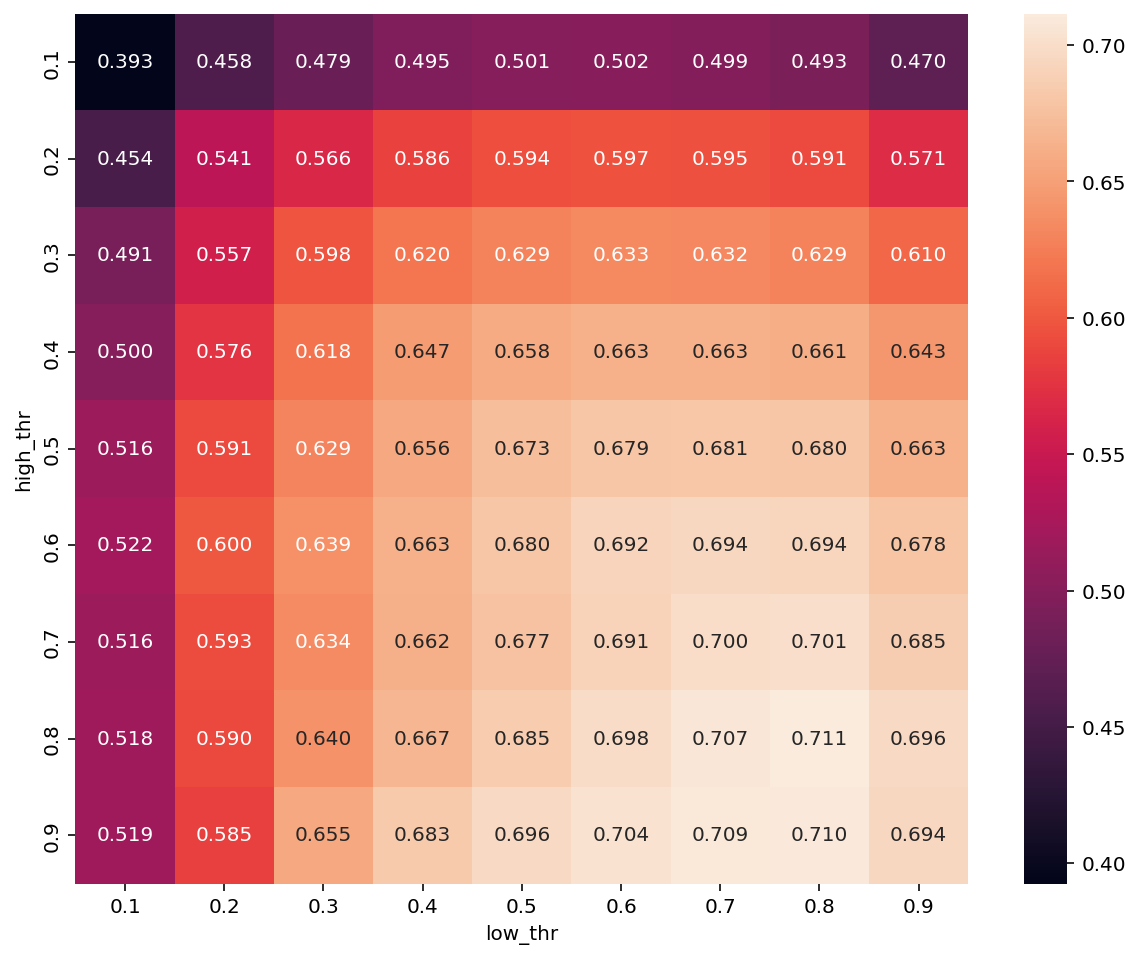}}
\caption{Silero MCC vs hysteresis thresholds.}
\label{fig_mcc_silero}
\end{figure}

\begin{figure}[htbp]
\centerline{\includegraphics[width=0.9\columnwidth]{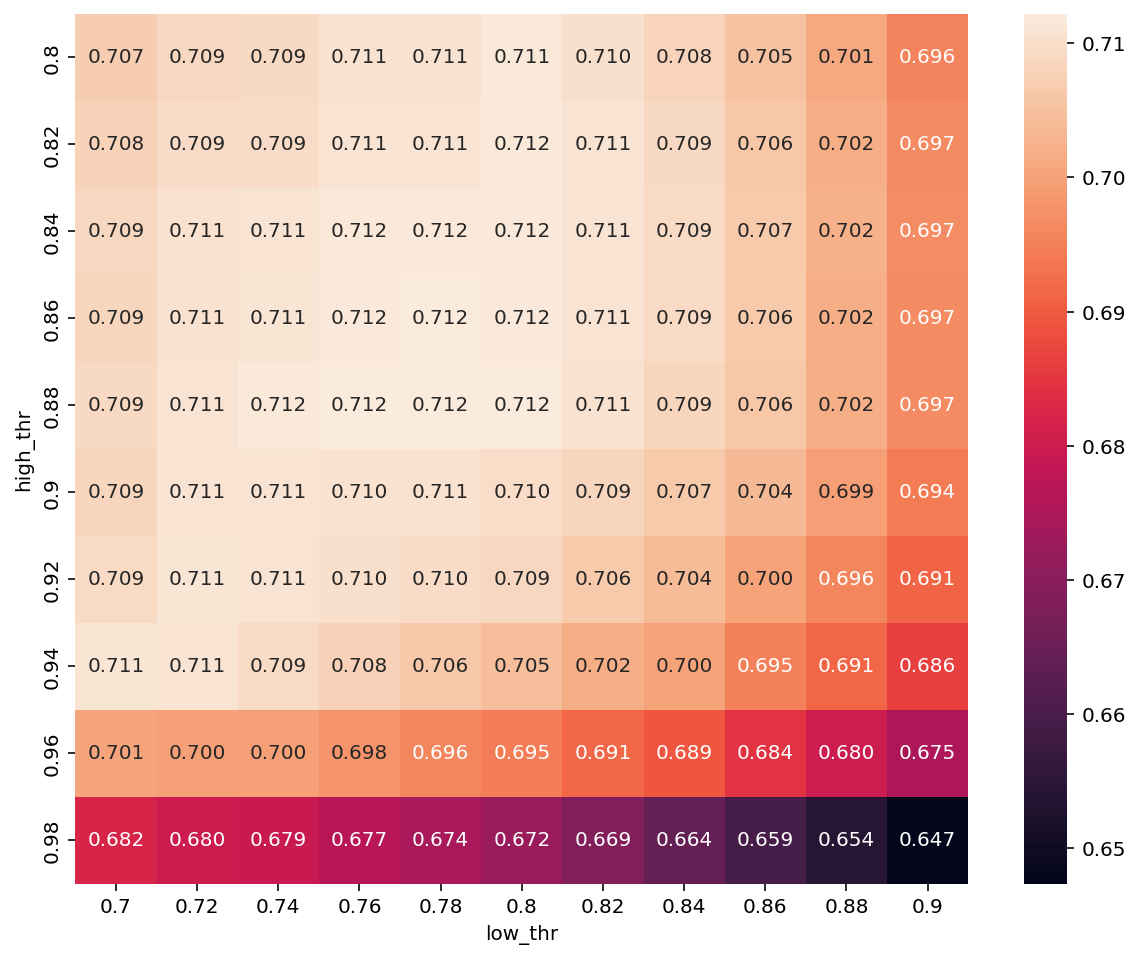}}
\caption{Silero MCC vs hysteresis thresholds (zoomed in around optimum).}
\label{fig_mcc_silero_zoomed}
\end{figure}

\begin{figure}[htbp]
\centerline{\includegraphics[width=0.9\columnwidth]{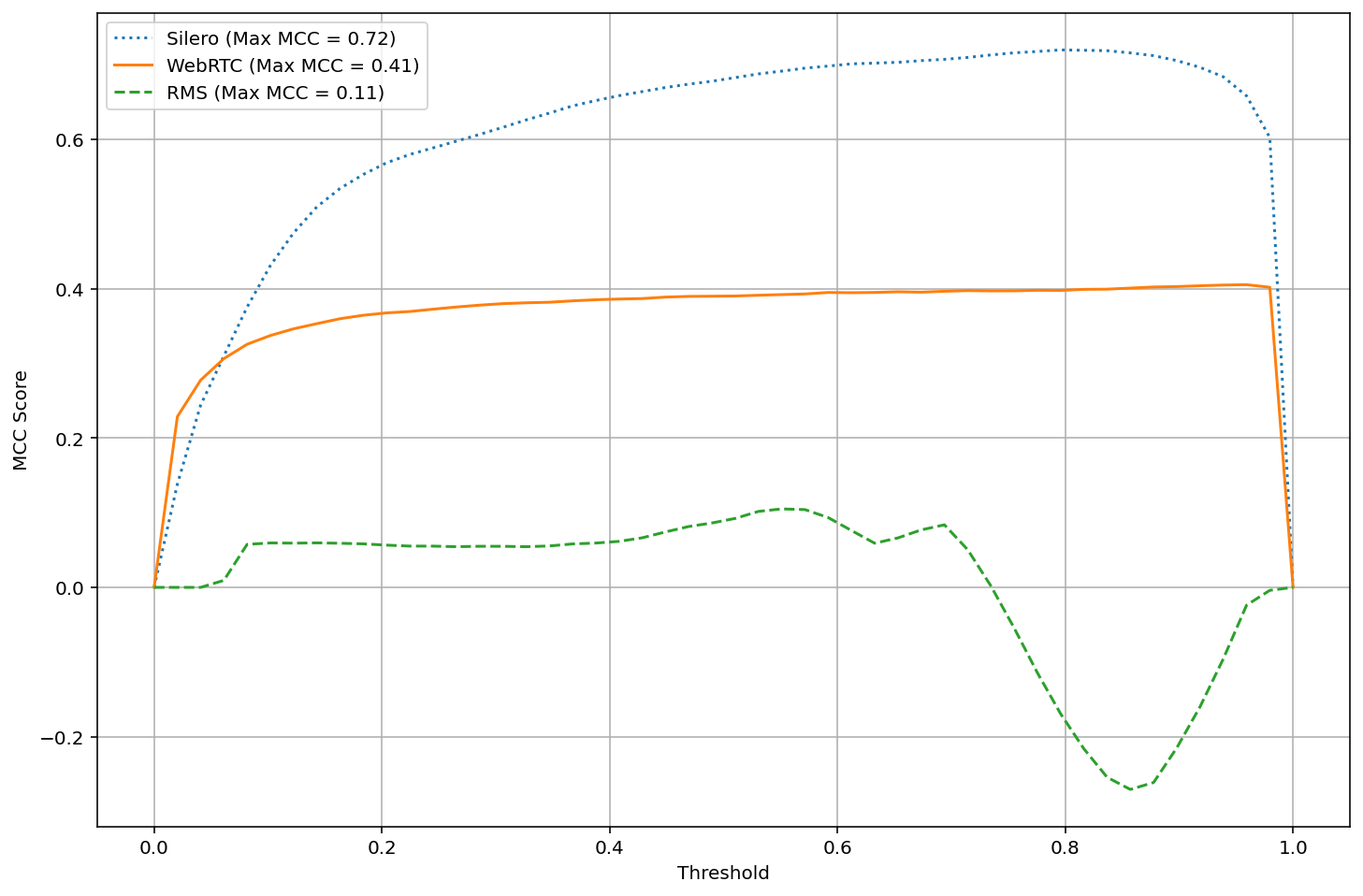}}
\caption{MCC vs decision threshold (window size 50ms).}
\label{fig_mcc_threshold}
\end{figure}

\begin{table}[htbp]
\centering
\caption{Maximum MCC scores for each VAD method with and without hysteresis post-processing.}
\label{tab:vad_comparison}
\begin{tabular}{lcc}
\toprule
\textbf{VAD Method} & \textbf{MCC} & \textbf{MCC with Hysteresis} \\
\midrule
RMS             & 0.11         & 0.10                       \\
WebRTC          & 0.41         & 0.47                       \\
Silero          & 0.72         & 0.71                       \\
\bottomrule
\end{tabular}
\end{table}

As seen in Table \ref{tab:vad_comparison}, splitting the threshold into two with hysteresis provided no significant benefit for RMS and Silero but did provide a slight benefit for WebRTC.

\FloatBarrier  
\section{Limitations}
\begin{itemize}
    \item The dataset, while diverse, is small (1.1-hour) and not exhaustive.
    \item The benefits of hysteresis might be better explored against raw VAD windows. The simple approach taken here of averaging results across subwindows to get a result on a bigger window naturally introduces smoothing which may reduce the effect of adding hysteresis.
    \item The ground truth labeling of speech considers entire phrases of speech as a single speech label for the phrase's entire duration which may overestimate the benefit of larger windows because this incorrectly penalizes classifying the gaps between words as non-speech.
\end{itemize}

\section{Conclusion}
In this work, we investigated the tradeoff between analysis window size and classification accuracy for three voice activity detectors: RMS, WebRTC, and Silero. Our experiments demonstrated that larger window sizes made from averaging subwindows generally lead to a degradation in performance as measured by ROC AUC and Average Precision. The modern, neural-based Silero VAD consistently and significantly outperformed the widely-used WebRTC VAD, which in turn was superior to the simple RMS energy-based baseline. Furthermore, we found that applying hysteresis post-processing provides a notable improvement to WebRTC's performance, increasing its peak MCC score, while offering no significant benefit to Silero or RMS.

\end{document}